\documentclass[11pt]{article}
\usepackage{epsfig}
\begin{document}

\title{Discrete
$Z^{\gamma}$: embedded circle patterns with the combinatorics of
the square grid  and
 discrete Painlev\'e equations}

\author{{\Large
Agafonov S.I. } \\
\\    Department of Mathematical Sciences \\
Loughborough University \\
    Loughborough, Leicestershire LE11 3TU \\
United Kingdom \\
    e-mail: {\tt
S.Agafonov@lboro.ac.uk} }
\date{}
\maketitle

\unitlength=1mm

\newtheorem{theorem}{Theorem}
\newtheorem{proposition}{Proposition}
\newtheorem{lemma}{Lemma}
\newtheorem{corollary}{Corollary}
\newtheorem{definition}{Definition}

\pagestyle{plain}

\begin{abstract}
A discrete analogue of the holomorphic map $z^{\gamma }$ is
studied. It is given by  Schramm's circle pattern with the
combinatorics of the square grid. It is shown that the
corresponding
 circle patterns are
embedded and described by special separatrix solutions of discrete
Painlev\'e equations. Global properties of these solutions, as
well as of the discrete $z^{\gamma }$, are established.
\end{abstract}

\section{Introduction}

The theory of circle patterns  is a fast developing field of
research on the border of complex analysis and discrete geometry.
Recent progress in this area has its origin in Thurston's idea

\cite{T} about approximating the Riemann mapping by circle
packings. By now some aspects of the theory of circle patterns, as
discrete analogs of conformal mappings, are well understood, while
the others are still waiting to be clarified. Classical circle
packings comprised of disjoint open disks were later generalized
to circle patterns where the disks may overlap (see for example
\cite{H}). Different underlying combinatorics were considered.
Schramm introduced a class of circle patterns with the
combinatorics of the square grid \cite{Schramm};
 hexagonal
circle patterns with constant multi-ratios  were studied by
Bobenko, Hoffman and Suris in \cite{BHS}; a new rich class of
hexagonal patterns with constant intersection angles was
investigated in \cite{BH}.

The convergence of discrete conformal maps represented by circle
packings was proven by Rodin and Sullivan \cite{RS}. For a
prescribed regular combinatorics this result was refined. He and
Schramm \cite{HS} showed that  for hexagonal packings the
convergence is $C^{\infty}.$ The uniform convergence for circle
patterns with the combinatorics of the square grid and orthogonal
neighboring circles was established by Schramm \cite{Schramm}.

Other facts underlining the striking analogy between circle
patterns and the classical theory are the uniformization theorem
 concerning circle packing realizations of
cell complexes of a prescribed combinatorics \cite{BS}, discrete
maximum principle, Schwarz's lemma
  and rigidity properties \cite{MR},\cite{H}, discrete Dirichlet
principle \cite{Schramm}.

It turned out that an effective approach to the description and
the construction of
 circle patterns with
overlapping circles is given by the theory of integrable systems
(see \cite{BPD},\cite{BHS},\cite{BH}). In particular, Schramm's
circle patterns studied in this paper are governed by a difference
equation which is the stationary Hirota equation (see
\cite{Schramm}). This equation is an example of an integrable
difference equation. It appeared first in a different branch of
mathematics -- the theory of integrable systems (see \cite{Z} for
a good survey).

On the other hand, not very much is known about analogs of
standard holomorphic functions, although computer experiments give
evidence for their existence \cite{DS}. For circle packings with
 the hexagonal combinatorics the only explicitly
described examples are Doyle spirals, which are  discrete analogs
of exponential maps, \cite{Doy} and conformally symmetric
packings, which are analogs of a quotient of Airy functions
\cite{BHConf}. For patterns with overlapping circles more explicit
examples are known: discrete versions of ${\rm exp}(x)$,
 ${\rm erf}(x)$ \cite{Schramm},
$z^{\gamma}$, $ {\rm log}(x)$ \cite{AB} are constructed for
patterns with underlying combinatorics of the square grid;
$z^{\gamma}$, ${\rm log}(x)$ are also described for hexagonal
patterns
 with both multi-ratio \cite{BHS}
and constant angle \cite{BH} properties.

Whereas  computer experiments reveal a regular behavior of the
circle patterns corresponding to discrete
 $z^{\gamma}$ and ${\rm log}(x)$, only the local property of
immersionness  was proved for the Schramm's patterns  \cite{AB}.
 This
property turned out to be connected with special solutions of
discrete Painlev\'e II equations, thus giving geometrical
 interpretation thereof.
The aim of this paper is to prove the global property of
embeddedness for
 the square grid circle patterns corresponding to $z^{\gamma}$ (which
was conjectured in \cite{AB}).

\begin{figure}[th]
\begin{center}
\epsfig{file=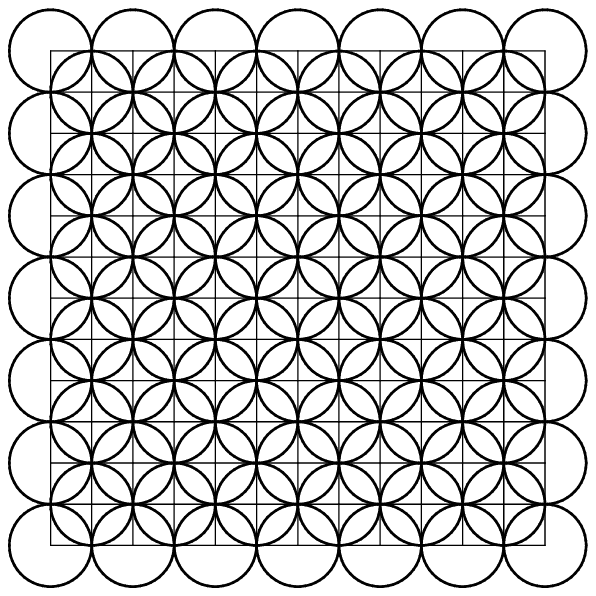,width=45mm}
\begin{picture}(20,50)
\put(2,22){\vector(1,0){15}} \put(7,27){ \it  \huge f}
\end{picture}
\epsfig{file=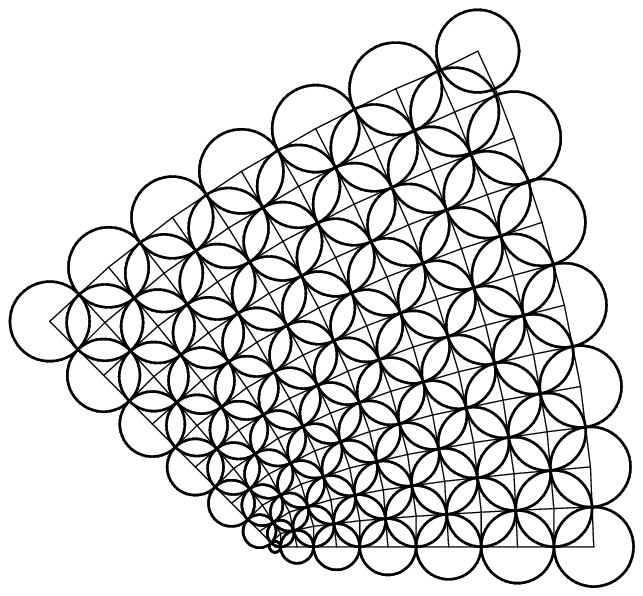,width=50mm} \caption{Schramm's circle
patterns as discrete conformal map.  Shown is the discrete version
of the holomorphic mapping $z^{3/2}$.} \label{CPMap}
\end{center}
\end{figure}

To visualize the analogy between Schramm's circle patterns and
conformal maps, consider regular patterns composed of unit circles
and suppose that the radii are being deformed so as to preserve
the orthogonality of neighboring circles and the tangency of
half-neighboring ones. Discrete maps taking intersection points of
the unit circles of the standard regular patterns to the
respective points of the deformed patterns mimic classical
holomorphic functions, the deformed radii being analogous to
$|f'(z)|$ (see Fig. \ref{CPMap}).

It is easy to show that the lattice comprised of the centers of
circles of  Schramm's pattern and their intersection

points is a special discrete conformal mapping (see Definition
\ref{DCM} below). The latter were introduced in \cite{BPdis} in
the frames of discrete integrable geometry, originally without any
relation to circle patterns.

\begin{figure}[th]
\begin{center}
\epsfig{file=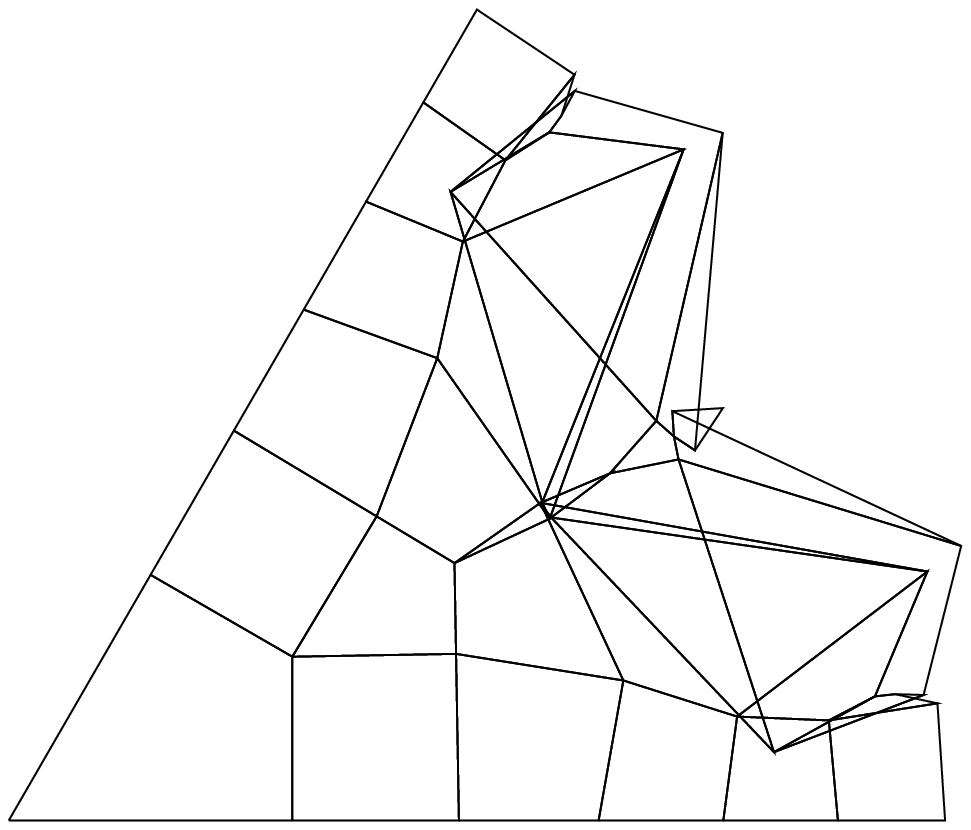,width=50mm}
\epsfig{file=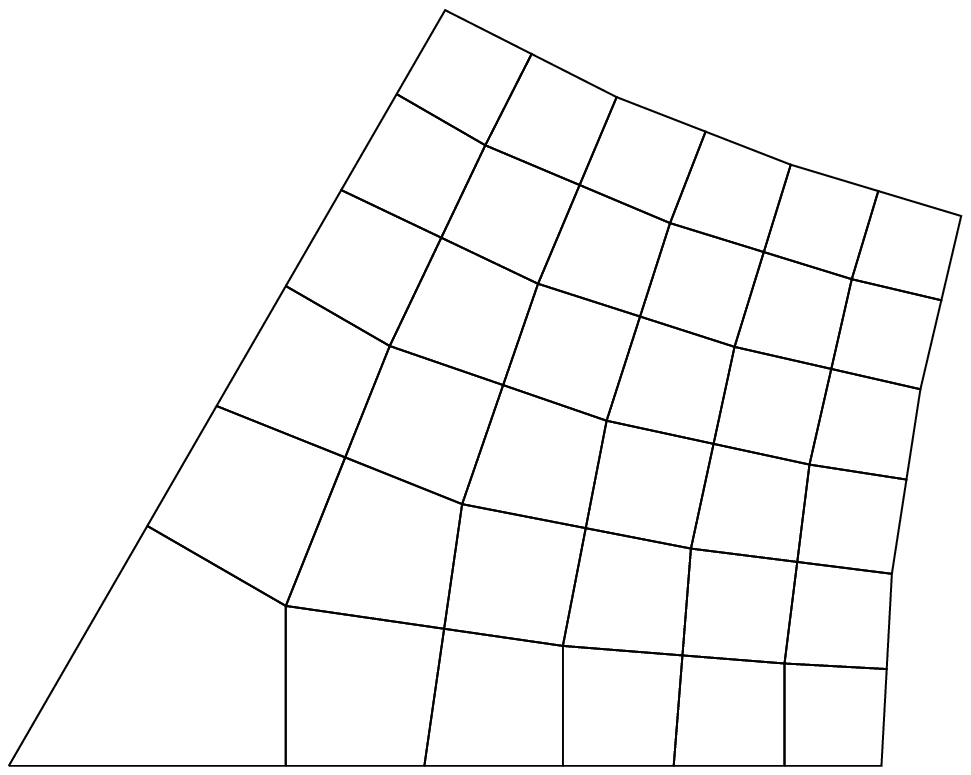,width=50mm} \caption{Two discrete
conformal maps with close initial data $n=0, m=0$. The second
lattice describes a discrete version of the holomorphic mapping
$z^{2/3}$.} \label{correct-wrong}
\end{center}
\end{figure}

\begin{figure}[t]
\begin{center}
\epsfig{file=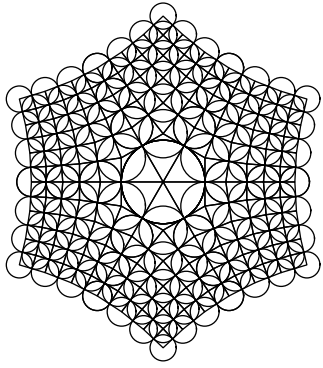, width=50mm}
\end{center}
\caption{Schramm's circle pattern corresponding to $Z^{2/ 3}$}
\label{f.2over3}
\end{figure}

\begin{definition} A map  $f\ : \ {\bf Z^2\ \rightarrow
\ {\bf R^2}={\bf C}}$ is called a discrete conformal map if all
its elementary quadrilaterals are conformal squares, i.e., their
cross-ratios are equal to -1: $$
q(f_{n,m},f_{n+1,m},f_{n+1,m+1},f_{n,m+1}):= $$
\begin{equation}
\frac{(f_{n,m}-f_{n+1,m})(f_{n+1,m+1}-f_{n,m+1})}
{(f_{n+1,m}-f_{n+1,m+1})(f_{n,m+1}-f_{n,m})}=-1 \label{q}
\end{equation}
\label{DCM}
\end{definition}
This definition is motivated by the following properties:\\ 1) it
is M\"obius invariant,\\ 2) a smooth map $f:\ D\ \subset {\bf C}
\to {\bf C}$ is conformal (holomorphic or antiholomorphic) if and
only if $$ \lim _{\epsilon \to 0}q(f(x,y),f(x+\epsilon
,y)f(x+\epsilon ,y+\epsilon )f(x,y+\epsilon ))=-1 $$ for all
$(x,y)\in D$. For some examples see \cite{BPdis},\cite{TH}. A
naive method to construct a discrete analogue of the function
$f(z)=z^{\gamma }$ is to start with $f_{n,0}=n^{\gamma }, \ n\ge
0$, $f_{0,m}=(im)^{\gamma }, \ m\ge 0$ and then compute $f_{n,m}$
for any $n,m > 0$ using equation (\ref{q}). But so determined map
has a behavior which is far from that of usual holomorphic maps.
Different elementary quadrilaterals  overlap (see the left lattice
in Fig. \ref{correct-wrong}).
\begin{definition}
A discrete conformal map $f_{n,m}$ is called embedded if inner
parts of different elementary quadrilaterals
$(f_{n,m},f_{n+1,m},f_{n+1,m+1},f_{n,m+1})$ do not intersect.
\end{definition}
\noindent The condition for discrete conformal map to be embedded
can be relaxed as follows.
\begin{definition}
A discrete conformal map $f_{n,m}$ is called an immersion if inner
parts of adjacent elementary quadrilaterals
$(f_{n,m},f_{n+1,m},f_{n+1,m+1},f_{n,m+1})$ are disjoint.
\end{definition}
To construct an embedded discrete analogue of $z^{\gamma },$ which
is the right lattice presented in Fig. \ref{correct-wrong}, a more
complicated approach is needed. Equation (\ref{q}) can be
supplemented with the following nonautonomous constraint:
\begin{equation}
\gamma f_{n,m}=2n\frac{(f_{n+1,m}-f_{n,m})(f_{n,m}-f_{n-1,m})}
{(f_{n+1,m}-f_{n-1,m})}+2m\frac{(f_{n,m+1}-f_{n,m})(f_{n,m}-f_{n,m-1})}
{(f_{n,m+1}-f_{n,m-1})}, \label{c}
\end{equation}
which plays a crucial role in this paper. This constraint, as well
as its compatibility with (\ref{q}), is derived from some
monodromy problem (see \cite{AB}). Let us assume $0<\gamma < 2$
and denote ${\bf Z^2_{+}}=\{ (n,m) \in {\bf Z^2}: n,m \ge 0 \}.$
Motivated by the asymptotics of the constraint (\ref{c}) at $n,m
\rightarrow \infty$ and the properties $$ z^\gamma({\bf R_+}) \in
{\bf R_+}, \ \ z^{\gamma}(i{\bf R_+}) \in e^{\gamma \pi i/2}{\bf
R_+} $$ of the holomorphic mapping $z^{\gamma}$ we use the
following definition \cite{BPD} of the "discrete" $z^{\gamma}$.
\begin{definition}
The discrete conformal map $Z^{\gamma}\ : \ {\bf Z^2_+\
\rightarrow \ {\bf C}}, \  0<\gamma < 2\ $ is the solution of
(\ref{q}),(\ref{c}) with the initial conditions
\begin{equation}
Z^{\gamma}(0,0)=0, \ \ Z^{\gamma}(1,0)=1, \ \
Z^{\gamma}(0,1)=e^{\gamma \pi i/2}. \label{initial}
\end{equation}
\label{def}
\end{definition}
Obviously, $Z^{\gamma}(n,0)\in {\bf R_{+}}$ and
$Z^{\gamma}(0,m)\in e^{\gamma \pi i/2} ({\bf R_{+}})$  for any  $
n,m \in {\bf N}$ .\\
 Fig. \ref{correct-wrong} suggests that $Z^{\gamma }$ is
embedded. The corresponding theorem is the main result of this
paper.
\begin{theorem}
The discrete map $Z^{\gamma }$ for $0<\gamma <2$ is embedded.
\label{main}
\end{theorem}
The proof is based on the analysis of geometric and algebraic
properties of the corresponding lattices. In Section 2 a brief
review of the necessary results from \cite{AB} is given. It is
shown that $Z^{\gamma }$ corresponds to a circle pattern of
Schramm's type. (The circle pattern corresponding to $Z^{2/3}$ is
presented in Fig. \ref{f.2over3} )
Next, analyzing the equations for radii of the circles we show
that in order to prove that $Z^{\gamma }$ is embedded it is enough
to establish a special property of the separatrix solutions
$P_{N,M},~Q_{N,M}$
 of
the following infinite sequence of systems of ordinary difference
equations of Painlev\'e type ($N$ labels  the system): $$
Q_{N,M+1}=\frac{
(M+N)P_{N,M}(P_{N,M}-Q_{N,M}^2)-(M-N)Q_{N,M}^2(1+P_{N,M})
}{Q_{N,M}( (M+N)(Q_{N,M}^2-P_{N,M})-(M-N)P_{N,M}(1+P_{N,M}) )} $$
$$ P_{N,M+1}=\frac{ (2M+\gamma )P_{N,M} + (2N+\gamma
)Q_{N,M}Q_{N,M+1}   }{ (2(N+1)-\gamma )P_{N,M} + (2(M+1)-\gamma
)Q_{N,M}Q_{N,M+1}  } $$ Namely, it is shown that $Z^{\gamma }$ is
embedded if the infinite sequence of solutions $Q_{N,M},P_{N,M}$
 with special initial data $Q_{N,N},P_{N,N}$
is subject to $$ (\gamma-1)(Q_{N,M}^2 - P_{N,M})\ge 0, \
Q_{N,M}>0, \ P_{N,M}>0 $$ The existence and uniqueness of the
corresponding initial data are given in Section 3.

For $N=0$ the system for $Q_{N,M},P_{N,M}$  reduces to the special
case of discrete Painlev\'e equation dPII
\begin{equation}
\label{dPII}
(n+1)(x_n^2-1)\left(\frac{x_{n+1}-ix_n}{i+x_nx_{n+1}}\right)-
 n(x_n^2+1)\left(\frac{x_{n-1}+ix_n}{i+x_{n-1}x_n}\right) =
\gamma x_n
\end{equation}
(see \cite{FIK}, \cite{NRGO} for more examples). In \cite{AB} this
equation was the main tool to prove the immersion of $Z^{\gamma}$:
 it was shown that $Z^{\gamma }$ is immersed if
the unitary solution $x_n=e^{i\alpha_n}$ of this equation with
$x_0=e^{i\gamma \pi/4}$ lies in the sector $0<\alpha _n < \pi/2.$

\begin{figure}[ht]
\begin{center}
\epsfig{file=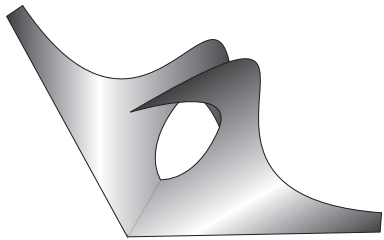, width=50mm}
\end{center}
\caption{Surface glued of quadrilaterals of immersed but
non-embedded discrete map. } \label{surface}
\end{figure}

Similar problems have been studied in the frames of the
isomonodromic deformation method \cite{IN}. In particular,
connection formulas were derived. These formulas describe the
asymptotics of solutions  for $n \to \infty$ as a function of
initial conditions (see in particular \cite{FIK}). These methods
seem to be insufficient for our purposes since we need to control
the behavior of solutions for finite $n$'s  as well. The geometric
origin of our equations permits us to prove the abovementioned
properties by purely geometric methods. To illustrate the
difference between the immersed and embedded discrete conformal
maps, let us imagine that the elementary quadrilaterals of the map
are made of elastic inextensible material and glued along the
corresponding edges to produce
 a
surface with a border. If this surface is immersed it is locally
flat. Being dropped down it will not have folds. At first sight it
seems to be sufficient to give embeddedness, provided $Z_{n,0}\to
\infty$ and $Z_{0,m}\to \infty$ as $n \to \infty$ (which follows
from
$
Z^{\gamma }_{n,0}=\frac{2c(\gamma )}{\gamma}\left(\frac {n}{2}
\right)^{\gamma }\left(1+O\left(\frac{1}{n^2} \right) \right), \ \
n \to \infty,
$
see \cite{AB}). But a surface with such properties still may  have
some limit curve with self-intersections thus giving overlapping
quadrilaterals. Hypothetical example of such a surface is shown in
Fig. \ref{surface}.
\section{Circle
patterns and $Z^{\gamma}$} Let us recall the necessary results
from \cite{AB}. $Z^{\gamma}$ of Definition \ref{def} determines a
special case of circle patters with the combinatorics of the
square grid as defined by Schramm in \cite{Schramm}. Indeed,
discrete $f_{n,m}$ satisfying (\ref{q}) and (\ref{c}) with initial
data $f_{0,0}=0$, $f_{1,0}=1$,  $f_{0,1}=e^{i \alpha }$   has the
equidistant property $$ f_{2n,0}-f_{2n-1,0}=f_{2n+1,0}-f_{2n,0},\
\ f_{0,2m}-f_{0,2m-1}=f_{0,2m+1}-f_{0,2m} $$ for any $n\ge1$,
$m\ge1$. \label{equi} \noindent Given initial $f_{0,0}$, $f_{0,1}$
and $f_{1,0}$ the constraint (\ref{c}) allows one to compute
$f_{n,0}$ and $f_{0,m}$ for all $n,m \ge 1.$ Now using equation
(\ref{q}) one can successively compute $f_{n,m}$ for any $n,m \in
{\bf N}$. Observe that if
$|f_{n+1,m}-f_{n,m}|=|f_{n,m+1}-f_{n,m}|$
 then the quadrilateral
$(f_{n,m},f_{n+1,m},f_{n+1,m+1},f_{n,m+1})$ is of the kite form --
it is inscribed in a circle and is symmetric with respect to the
diameter of the circle $[f_{n,m}, f_{n+1,m+1}].$ If  the angle at
the vertex $f_{n,m}$ is $\pi /2$ then the quadrilateral
 $(f_{n,m},f_{n+1,m},f_{n+1,m+1},f_{n,m+1})$ is of the kite form
too. In this case the quadrilateral is symmetric with respect to
its diagonal $[f_{n,m+1}, f_{n+1,m}]$.

\begin{proposition}\cite{AB}
Let $f_{n,m}$ satisfy (\ref{q}) and (\ref{c}) in ${\bf Z^2_+}$
with initial data $f_{0,0}=0$, $f_{1,0}=1$,  $f_{0,1}=e^{i \alpha
}.$ Then all the elementary quadrilaterals $(f_{n,m},
f_{n+1,m},f_{n+1,m+1},f_{n,m+1})$ are of the kite form. All edges
at the vertex $f_{n,m}$ with $n+m=0 \ ({\rm mod} \ 2)$ are of the
same length $$
|f_{n+1,m}-f_{n,m}|=|f_{n,m+1}-f_{n,m}|=|f_{n-1,m}-f_{n,m}|=|f_{n,m-1}-f_{n,m}|.
$$ All angles between the neighboring edges at the vertex
$f_{n,m}$ with $n+m=1 \ ({\rm mod} \ 2)$  are equal to $\pi /2.$
\label{kite}
\end{proposition}
\noindent Proposition \ref{kite} implies that for any $n,m: \
n+m=0 \ ({\rm mod }\ 2)$ the points $f_{n+1,m},$ $f_{n,m+1},$
$f_{n-1,m},$ $f_{n,m-1}$ lie on a circle with the center
$f_{n,m}$.
\begin{corollary}\cite{AB}
The circumscribed circles of the quadrilaterals $(f_{n-1,m},
f_{n,m-1},f_{n+1,m},f_{n,m+1})$ with $n+m=0 \ ({\rm mod} \ 2)$
form a circle pattern of Schramm type (see \cite{Schramm}), i.e.
the circles of neighboring quadrilaterals intersect orthogonally
and the circles of
 half-neighboring
quadrilaterals with common vertex are tangent (see Fig. 3).
\end{corollary}

Consider the sublattice $\{n,m: \ n+m=0 \ ({\rm mod} \ 2)\}$ and
denote by $\bf V$ its quadrant $$ {\bf V}=\{z=N+iM:\ N,M \in {\bf
Z^2}, M \ge |N| \}, $$ where $$ N=(n-m)/2, \ \ M=(n+m)/2. $$
 We
will use complex labels $z=N+iM$ for this sublattice. Denote by
$C(z)$ the circle of the radius

\begin{equation}
R(z)=|f_{n,m}-f_{n+1,m}|=|f_{n,m}-f_{n,m+1}|=|f_{n,m}-f_{n-1,m}|=|f_{n,m}-f_{n,m-1}|
\label{rmap}
\end{equation}
with the center at $f_{N+M,M-N}=f_{n,m}.$ From Proposition
\ref{kite} it follows that any two circles $C(z)$, $C(z')$ with
$|z-z'|=1$ intersect orthogonally and any two circles $C(z)$,
$C(z')$ with $|z-z'|=\sqrt{2}$ are tangent.
\\
Let $\{C(z)\}, \ z\in {\bf V}$ be a  circle pattern of Schramm
type on the complex plane. Define $f_{n,m}: {\bf Z^2_+ \to C}$ as
follows:\\ a) if $n+m=0 \ ({\rm mod} \ 2)$ then $f_{n,m}$ is the
center of $C(\frac{n-m}{2}+i\frac{n+m}{2}),$\\ b) if $n+m=1 \
({\rm mod} \ 2)$ then
$
f_{n,m}:=C(\frac{n-m-1}{2}+i\frac{n+m-1}{2}) \cap
C(\frac{n-m+1}{2}+i\frac{n+m+1}{2})=
C(\frac{n-m+1}{2}+i\frac{n+m-1}{2}) \cap
C(\frac{n-m-1}{2}+i\frac{n+m+1}{2}).
$
Since all elementary quadrilaterals $(f_{n,m},
f_{n+1,m},f_{n+1,m+1},f_{n,m+1})$ are of the kite form equation
(\ref{q}) is satisfied automatically. In what follows the function
$f_{n,m},$ defined as above by a) and b) is called {\it a discrete
conformal map corresponding to the circle pattern $\{C(z)\}$ .}
\begin{theorem}\cite{AB}
Let $f_{n,m}$ satisfying (\ref{q}) and (\ref{c}) with initial data
$f_{0,0}=0$, $f_{1,0}=1$, $f_{0,1}=e^{i \alpha }$ be an immersion,
then $R(z)$ defined by (\ref{rmap}) satisfies the following
equations:
\begin{equation}
\begin{array}{l}
R(z)R(z+1)(-2M-\gamma )+R(z+1)R(z+1+i)(2(N+1)-\gamma )+\\ \qquad
R(z+1+i)R(z+i)(2(M+1)-\gamma )+R(z+i)R(z)(-2N-\gamma )=0,
\end{array}
\label{square}
\end{equation}
for $z\in {\bf V}_l:={\bf V}\cup \{-N+i(N-1)|N\in {\bf N}\}$ and
\begin{equation}
\begin{array}{l}
(N+M)(R(z)^2-R(z+1)R(z-i))(R(z+i)+R(z+1))+\\ \qquad
(M-N)(R(z)^2-R(z+i)R(z+1))(R(z+1)+R(z-i))=0,
\end{array}
\label{Ri}
\end{equation}
for $z\in {\bf V}_{rint}:={\bf V}\backslash \{\pm N+iN|N\in {\bf
N}\}.$\\ Conversely let $R(z): {\bf V} \to {\bf R_+}$ satisfy
(\ref{square}) for $z\in {\bf V}_l$ and (\ref{Ri}) for $z\in {\bf
V}_{rint}.$ Then $R(z)$ define an immersed circle packing with the
combinatorics of the square grid. The corresponding discrete
conformal map $f_{n,m}$ is an immersion and satisfies (\ref{c}).
\label{eqforR}
\end{theorem}
\noindent From the initial condition (\ref{initial}) we have
\begin{equation}
R(0)=1, \ \ \ R(i)=\tan \frac{\gamma \pi}{4}. \label{Rinitial}
\end{equation}
Equation (\ref{square}) at $z=N+iN$ and $z=-N+i(N-1),\ N\in {\bf
N}$ reads as
\begin{equation}
R(\pm (N+1)+i(N+1))=\frac{2N+\gamma}{2(N+1)-\gamma }R(\pm N+iN)
\label{recR}
\end{equation}
and defines $R(\pm N+iN)$ for all $N \in {\bf N}.$ Now equations
(\ref{square},\ref{Ri}) determine $R(z)$ for any other $z\in {\bf
V}.$ Besides $R(z)$ satisfy
 (\ref{Ri}) at
$z=-N+iN, \ N \in {\bf N}$ which reads
 $$
 R(-N+1+iN)R(-N+i(N+1))=R^2(-N+iN).
 \label{line}
 $$
By symmetry one gets $$ R(N-1+iN)R(N+i(N+1))=R^2(N+iN). $$
 This equation
allows to compute $R(N+i(N+1))$. (Moreover, it implies that the
center $O$ of $C(N+iN)$ and the points $A=C(N+iN) \cap C(N-1+iN)
\backslash C(N-1+iN) \cap C(N+i(N+1))$
 and
$B=C(N+iN) \cap C(N+i(N+1)) \backslash C(N-1+iN) \cap C(N+i(N+1))$
are collinear so the points $f_{n,0}$ lie on a straight line.) To
prove Theorem \ref{main} we consider more general initial
condition:
\begin{equation}
R(0)=1, \ \ \ R(i)=a\tan \frac{\gamma \pi}{4}. \label{Rainitial}
\end{equation}
Now the radii $R(N+iN),R(N+i(N+1)$ can be represented  in terms of
$\Gamma$-~function:
\begin{equation}
R(N+iN)=c(\gamma )\frac{\Gamma(N+\gamma/2)}{\Gamma(N+1-\gamma
/2)}, \ \
  \mbox{where   }
c(\gamma )=\frac{\gamma \Gamma(1-\gamma /2)}{2\Gamma(1+\gamma
/2)}, \label{Rline}
\end{equation}
\begin{equation}
R(N+i(N+1))=\left( a\tan \frac {\gamma \pi}{4} \right)^{(-1)^N}
\left(\frac{ (2(N-1)+\gamma )(2(N-3)+\gamma )(2(N-5)+\gamma )...}
{(2N-\gamma )(2(N-2)-\gamma )(2(N-4)-\gamma )... } \right)^2.
\label{Ronline}
\end{equation}
Theorem \ref{eqforR} allows us to reformulate the property of the
circle lattice to be immersed completely in terms of the system
(\ref{square},\ref{Ri}). Namely to prove that $Z^{\gamma}$ is an
immersion one should show that the solution of the system
(\ref{square},\ref{Ri}) with initial data (\ref{Rinitial}) is
positive for all $z \in {\bf V}$ (see \cite{AB}). To prove Theorem
\ref{main} one need more subtle property of this solution.
\begin{theorem}
If for a  solution R(z) of (\ref{square}, \ref{Ri}) with $\gamma
\ne 1$ and initial conditions (\ref{Rinitial})  holds
\begin{equation}
 R(z)>0, \
\ (\gamma -1)(R(z)^2 - R(z-i)R(z+1))\ge 0 \label{sign}
\end{equation}
in ${\bf V}_{int}$, then the corresponding discrete conformal map
is embedded. \label{convex}
\end{theorem}
\noindent {\it Proof:} Since $R(z)>0$ the corresponding discrete
conformal map is immersion (see \cite{Schramm,AB}).
  Consider
piecewise linear curve $\Gamma _n$ formed by segments
$[f_{n,m},f_{n,m+1}]$ where $n>0$ and $0\le m \le n-1$ and the
vector ${\bf v}_n(m)={\bf (f_{n,m}f_{n,m+1})}$ along this curve.
Due to Proposition \ref{kite} this vector
 rotates only in vertices with
$n+m=0 \ ({\rm mod} \ 2)$ as $m$ increases along the curve. The
sign of the rotation angle $\theta _n(m)$, where $-\pi < \theta
_n(m) < \pi$, $0<m<n$ is defined by the sign of expression
$R(z)^2-R(z+i)R(z+1)$  (note that there is no rotation if this
expression vanishes), where
 $z=(n-m)/2+i(n+m)/2$ is a label
for the circle with the center in $f_{n,m}$. If $n+m=1\ ({\rm mod
\ 2})$ define $\theta _n(m)=0$. Now the theorem hypothesis and
equation (\ref{Ri}) imply that
 the vector
${\bf v}_n(m)$ rotates with increasing $m$ in the same direction
for all $n$, and namely, clockwise for $\gamma <1$ and
counterclockwise for $\gamma >1$. Consider the sector $B:=\{
z=re^{i\varphi}:r\ge 0,\ 0\le \varphi \le \gamma \pi /4 \}$. The
terminal
 points of the curves $\Gamma _n$ lie
on the sector border.
\begin{lemma}
For the curve $\Gamma _n$ holds:
\begin{equation}
\left |\sum_{m=1}^{n-1} \theta _n(m) \right | < \frac{
\pi}{4}(1+|1-\gamma |) \label{ineq}
\end{equation}
\label{rot}
\end{lemma}
\noindent {\it Proof of Lemma \ref{rot}}: Let us prove the
inequality (\ref{ineq}) for $1 < \gamma <2$ by induction for $n$.
For $n=1$ the inequality is obviously true since the curve $\Gamma
_1$ is a segment perpendicular to $\bf R_+$.
 Define the angle $\alpha _n(m)$ between
$i{\bf R_+}$ and the vector ${\bf v}_n(m)$
 by
$f_{n,m+1}-f_{n,m}=e^{i(\alpha _n(m)+\pi/2)}|f_{n,m+1}-f_{n,m}|$,
where $0 \le \alpha _n(m) <2\pi$, $0\le m <n$. Then
$\sum_{m=1}^{l} \theta _n(m)=\alpha _n(l)-\alpha _n(0)+2\pi
k_n(l)$ for some positive integer $k_n(l)$ increasing with $l$.
Note that $\alpha _n(0) < \pi/2 $, which easily follows from
(\ref{Rline},\ref{Ronline})), and $\alpha _n(n-1) < (\frac{\gamma
\pi}{4}+\frac{\pi}{2})-\frac{\pi}{2}= \frac{\gamma \pi}{4} $ since
for immersed $Z^{\gamma }$ the angle between
 the vector ${\bf v}_n(n-1)$ and
$ e^{i\gamma \pi/4}{\bf R_+}$ is less then $\frac{\pi}{2}$. Let
(\ref{ineq}) holds for $n>1$: $\left |\sum_{m=1}^{n-1} \theta
_n(m) \right |=\sum_{m=1}^{n-1} \theta _n(m)  < \frac{\gamma
\pi}{4}$ (all $\theta _n(m)$ are positive for $1 < \gamma <2$).
That implies $k_n(l)$=0, since $k_n(l)=(\sum_{m=1}^{l} \theta
_n(m)-\alpha _n(l)+\alpha _n(0))/2\pi \le (\sum_{m=1}^{l} \theta
_n(m)+|\alpha _n(l)|+|\alpha _n(0)|)/2\pi  < (\gamma \pi /4
+\gamma \pi /4+\pi /2)/2\pi < 1$ and $k_n(l)$ is integer. Let
$\alpha _{n+1}(l)=\alpha _n(l)+\sigma _n(l)$. All elementary
quadrilaterals are of the kite form therefore  $|\sigma
_n(l)|<\pi/2$. Let us prove, that $k_{n+1}(m)=0$ for $0\le m \le
n+1$. Obviously $k_{n+1}(0)=0$. Assume $k_{n+1}(l)=0$ but
$k_{n+1}(l+1)>0$. The increment of l.h.s. of $$ \sum_{m=1}^{l}
\theta _{n+1}(m)=\alpha _{n+1}(l)-\alpha _{n+1}(0)+2\pi k_{n+1}(l)
$$ as $l\to l+1$ is $\theta _{n+1}(l+1)<\pi$. The increment of
r.h.s. is no less than $2\pi +\alpha _{n+1}(l+1)-\alpha _{n+1}(l)
\ge 2\pi-\alpha _{n+1}(l) \ge 2\pi -\alpha _n(l)-|\sigma _n(l)|
> 2\pi -\gamma \pi /4 -\pi/2 > \pi.$ The
obtained contradiction gives $k_{n+1}(l)=0$  and $\sum_{m=1}^{n}
\theta _{n+1}(m)=\alpha _{n+1}(n)-\alpha _{n+1}(0)
 \le \alpha _{n+1}(n) <
\frac{\gamma \pi}{4}.$  Lemma \ref{rot} is proved.

\smallskip
The obvious corollary of Lemma \ref{rot} is that the curve $\Gamma
_n$ has no self-intersection and lies in the sector $B$
 since the rotation of
the vector ${\bf v}_n(m)$ along the curve is less then $\gamma
\pi/4<\pi/2$. Each such curve cuts the sector $B$ into a finite
part and an infinite part. Since the curve $\Gamma _n$ is convex
and  the borders of all elementary quadrilaterals
$(f_{n,m},f_{n+1,m},f_{n+1,m+1},f_{n,m+1})$ for embedded
$Z^{\gamma }$ have the positive orientation
 the
segments of the curve $\Gamma _{n+1}$ lie in the infinite part.
Now the induction in $n$ completes the proof of Theorem
\ref{convex} for $1<\gamma <2$. The proof for
 $0<\gamma <1$ is similar. The differences
are that $\theta _n(m)$ is not positive, the angle $\alpha $ is
naturally defined as negative: $-2\pi <\alpha _n(m) <0$, so that
 $-\pi/2
<\alpha _n(0) \le 0 $ and $\frac{\gamma \pi}{4}(2-\gamma)  <\alpha
_n(n-1) < 0$. Details are left to the reader.
\begin{corollary}
$$ \lim _{n\to \infty}Z^{\gamma} _{n,m}=\infty, \ \ \lim _{m\to
\infty}Z^{\gamma} _{n,m}=\infty. $$
\end{corollary}
Since the terminal
 points of the
curves $\Gamma _n$ lie on the sector border the proof easily
follows from convexity of the curves $\Gamma _n$, inequality
(\ref{ineq})  and from $$ \lim _{n\to \infty}Z^{\gamma}
_{n,0}=\infty. $$

\section{
$Z^{\gamma}$ and discrete Painlev\'e equations} Let $R(z)$ be a
solution of (\ref{square},\ref{Ri}) with initial condition
(\ref{Rinitial}). For $z\in {\bf V}_{int}$ define
$P_{N,M}=P(z)=\frac{R(z+1)}{R(z-i)}$,
$Q_{N,M}=Q(z)=\frac{R(z)}{R(z-i)}$. Then equations
(\ref{square},\ref{Ri}) are rewritten as follows
\begin{equation}
Q_{N,M+1}=\frac{
(M+N)P_{N,M}(P_{N,M}-Q_{N,M}^2)-(M-N)Q_{N,M}^2(1+P_{N,M})
}{Q_{N,M}( (M+N)(Q_{N,M}^2-P_{N,M})-(M-N)P_{N,M}(1+P_{N,M}) )}
\label{QPainleve}
\end{equation}
\begin{equation}
P_{N,M+1}=\frac{ (2M+\gamma )P_{N,M} + (2N+\gamma
)Q_{N,M}Q_{N,M+1}   }{ (2(N+1)-\gamma )P_{N,M} + (2(M+1)-\gamma
)Q_{N,M}Q_{N,M+1}  } \label{PPainleve}
\end{equation}
The property (\ref{sign}) for (\ref{QPainleve},\ref{PPainleve})
reads as
\begin{equation}
(\gamma-1)(Q_{N,M}^2 - P_{N,M})\ge 0, \ Q_{N,M}>0, \ P_{N,M}>0.
\label{PQsign}
\end{equation}
Equations (\ref{QPainleve},\ref{PPainleve}) can be considered as a
dynamical  system for  variable $M$,
 the expressions (\ref{Rline},\ref{Ronline}) defining
initial conditions $P_{N,N+1},Q_{N,N+1}$ for
(\ref{QPainleve},\ref{PPainleve})
 as
functions of $a,N$, only $Q_{N,N+1}$ being dependent on $a$.
\begin{theorem}
There exists such $a>0$ that for  the solutions $R(z)$ of
(\ref{square},\ref{Ri}) with initial conditions (\ref{Rainitial})
holds (\ref{sign}). \label{existence}
\end{theorem}
\noindent {\it Proof:} Due to the following Lemma it is sufficient
to prove (\ref{sign}) only for $0<\gamma<1$.
\begin{lemma}
If $R(z)$ is a solution of (\ref{square},\ref{Ri}) for $\gamma$
then $1/R(z)$ is a solution of (\ref{square},\ref{Ri}) for $\tilde
\gamma =2-\gamma$.
\end{lemma}
Lemma is proved by straightforward computation.

\smallskip
Let $0<\gamma <1$ and $(P_{N,M},Q_{N,M})$ correspond to the
solution of (\ref{square},\ref{Ri}) defined by initial conditions
(\ref{Rainitial}). Define the real function $F(P)$ for $P\in {\bf
R_+}$:
 $$F(P)=\sqrt{P} \ {\rm for}
\
0\le P \le 1, \ F(P)=1\ {\rm for} \ 1\le P. $$ Designate $$
D_{u}:=\{(P,Q): P > 0, Q > F(P) \},\ \ D_{d}:=\{(P,Q): Q < 0 \},
$$ $$
 D_{0}:=\{(P,Q): P > 0, 0 \le
Q \le F(P) \}, \ \ D_f:=\{ (P,Q): P\le 0, Q\ge 0  \}. $$ Now
define the infinite sequences $\{q_n\},\{p_n \},\ n\in {\bf N}$ as
follows: $$ \{q_n(a)\}:=\{ Q_{0,1},Q_{0,2},
Q_{1,2},Q_{0,3},Q_{1,3},Q_{2,3},...,Q_{0,M},Q_{1,M},...,Q_{M-1,M},...
\}, $$ $$ \{p_n(a)\}:=\{ P_{0,1},P_{0,2},
P_{1,2},P_{0,3},P_{1,3},P_{2,3},...,P_{0,M},P_{1,M},...,P_{M-1,M},...
\}. $$ and the sets $$ A_u(n):=\{a\in {\bf R_+}:
(p_n(a),q_n(a))\in D_u, (p_k(a),q_k(a))\in D_0 \ \forall \ 0< k <n
\}, $$ $$ A_d(n):=\{a\in {\bf R_+}: (p_n(a),q_n(a))\in D_d,
(p_k(a),q_k(a))\in D_0 \ \forall \ 0< k <n \}. $$ $A_u(n)$ and
$A_d(n)$ are open sets since the denominators of
(\ref{QPainleve},\ref{PPainleve}) do not vanish in $D_0$. Indeed,
the curve $Q^2=P+\frac{(M-N)}{(M+N)}P(P+1)$ lies outside $D_0.$
Moreover, direct computation shows that
 $A_u(1)\ne \emptyset$ and $A_d(2)\ne
\emptyset,$ therefore the sets $$ A_u:=\cup A_u(k), \ \ \
A_d:=\cup A_d(k) $$ are not empty. Finally, define $$ A_0:=\{a\in
{\bf R_+}: (p_n(a),q_n(a))\in D_0,  \ \forall \ n \in{\bf N} \}.
$$ Note that $A_0,A_u,A_d$ are mutually disjoint and the sequences
$\{p_n\},\{q_n\}$ is so constructed that
\begin{equation}
\label{union} {\bf R_+}=A_0\cup A_u \cup A_d.
\end{equation}
Indeed $(P_{N,M},Q_{N,M})$ can not jump from $D_0$ into $D_f$ in
one step $M \to M+1$ since $P_{N,M+1}$ is positive for positive
 $P_{N,M},Q_{N,M},Q_{N,M+1}$.
The relation (\ref{union}) would be impossible for
$A_0=\emptyset$, since the connected set ${\bf R_+}$ can not be
covered by two open
 disjoint nonempty
subsets $A_u$ and $A_d$, therefore $A_0\ne \emptyset.$ Q.E.D.

\begin{proposition}
\label{true_a} The set $A_0$ consists of only one element, namely,
$A_0=\{1\}$.
\end{proposition}

\noindent {\it Proof:} A positive solution $R(z)$ of
(\ref{square},\ref{Ri})
 provides an immersed  discrete conformal map corresponding
to the circle patterns
 with radii $R(z)$. It was proven in \cite{AB}
that the initial value (\ref{Rainitial}) for immersed $Z^{\gamma}$
consists of only one element $a=1$. Q.E.D.

\medskip

\noindent {\it Proof of Theorem \ref{main}:} combining Theorems
\ref{convex} and \ref{existence} with Proposition \ref{true_a} one
easily deduces Theorem \ref{main}.

\section{Concluding remarks}

The approach suggested in this paper can be applied to prove the
embeddedness of the circle patterns corresponding to discrete
$Z^2$ and ${\rm Log}$. (These patterns were proved to be immersed
\cite{AB}.) Slightly modified, it seems to
 be also applicable to show global properties of the discrete $Z^K$
proposed in \cite{AB} for natural $K>2$. Details will be discussed
elsewhere.

\section{Acknowledgements}

The author would like to thank A.Bobenko and Yu.Suris for useful
discussions. This research was supported by the EPSRC grant No
Gr/N30941.

\end{document}